\documentclass[12pt]{article}
\usepackage{epsfig,graphicx}         

\def\be{\begin{eqnarray}}
\def\ee{\end{eqnarray}}
\def\beq{\begin{eqnarray}}
\def\eeq{\end{eqnarray}}
\newcommand{\ep}{\varepsilon}

\newcommand{\ice}[1]{\relax}

\newcommand{\GeV}{{\rm GeV}}

\newcommand{\nn}{\nonumber}

\newcommand{\bdbdbar}{$B_d^0$--${\bar B_d^0}$}
\begin{document}

\begin{center}
{\Large \bf Computing non-factorizable pQCD corrections to hadronic 
$B^0 - \bar B^0$ mixing matrix element within sum rules technique
for three point Green functions\footnote{Talk given at 
13th International Seminar on High Energy Physics "Quarks-2004", 
Pushkinskie Gory, Russia, May 24-30, 2004; supported in part by 
the Russian Fund for Basic Research under contract 03-02-17177
and INTAS grant under contract 03-51-4007.
To be published in the Proceedings.} 
}
\vskip 1cm
{\large \bf A. A.~Pivovarov}

{\em Institute for Nuclear Research of the Russian
Academy of Sciences, Moscow, 117312 Russia}
\end{center}
\vskip 2cm 

\begin{abstract}
\noindent In this talk I report on the results of a recent 
calculation of the $\alpha_s$ corrections to a three-point correlation
function at the three-loop 
level in QCD perturbation theory, which allows one to extract the matrix 
element of $B^0 -\bar B^0$ mixing with next-to-leading order
accuracy~\cite{main1}. The evaluation of mixing parameter
at NLO allows for a consistent analysis of $B^0 -\bar B^0$ mixing 
since the coefficient functions of the effective 
Hamiltonian for this process are known with the necessary accuracy. 
\end{abstract}

\noindent Presently the pattern of CP violation in the Standard Model
is under thorough experimental study in dedicated experiments by 
BABAR collaboration at SLAC~(e.g.~\cite{Aubert:2002rg,Aubert:2002ic})
and BELLE collaboration at KEK~(e.g.~\cite{Abe:2000yh,Abashian:2001pa}).
The standard mechanism of CP violation to be primarily tested is a 
Cabibbo-Kobayashi-Maskawa paradigm with mixing of (at least)
three quark generations~(for review, see e.g.~\cite{buhalla}). 
At the hadronic level the fact of quark mixing mainly reveals itself 
as mixing of neutral pseudoscalar mesons.
The most famous system where mixing occurs and has been studied in
much detail is the system of neutral kaons. 
Study of $K^0 - \bar K^0$ mixing strongly constrained 
the physics of heavy particles and allowed
to estimate the numerical value of the charm quark mass
from the requirement of GIM cancellation before the 
experimental discovery of charm~(see e.g.~\cite{kkreviewOLD}). 
At present the experimental studies of CP violation shifted to the
realm of heavy mesons for which they are considered more 
promising. In particular, recent experimental results for heavy charmed mesons 
$D (\bar u c)$ are encouraging~\cite{Petrov:2002is}. 
However the systems of $B_d (\bar d b)$ and $B_s (\bar s b)$ 
mesons are the most promising laboratory for performing 
a precision analysis of CP violation 
and mixing both experimentally and theoretically~\cite{reviewBB}. 

Mixing in any system of neutral pseudoscalar mesons 
is described by a 2x2 effective Hamiltonian or mass operator
$\left(M-i\Gamma/2\right)_{ij}$, $\{i,j\}=\{1,2\}$ where $M$ is
related to the mass spectrum of the system and $\Gamma$ describes the
widths of the mesons. In the presence of 
flavor violating interactions ($\Delta B=2$ in our particular case)
the effective Hamiltonian acquires non-diagonal terms. The difference 
between the values of the mass eigenstates of $B$ mesons 
$\Delta m = M_{heavy}-M_{light}\approx 2\left|M_{12}\right|$ is
precisely measured
$\Delta m= 0.489\pm0.005(stat)\pm 0.007(syst)~ps^{-1}$~\cite{PDG}.
With an accurate theoretical description of the mixing, it can be used 
to extract the top quark CKM parameters. 
In Wolfenstein's parametrization the CKM matrix elements reveal a 
hierarchy in magnitude.
In terms of Cabibbo angle $\lambda\approx 0.22$, this parametrization 
of the CKM matrix $V$ reads
\beq
V \simeq \left(\matrix{
 1-{1\over 2}\lambda^2 & \lambda
 & A\lambda^3 \left( \rho - i\eta \right) \cr
 -\lambda ( 1 + i A^2 \lambda^4 \eta )
& 1-{1\over 2}\lambda^2 & A\lambda^2 \cr
 A\lambda^3\left(1 - \rho - i \eta\right) & -A\lambda^2 
& 1\cr}\right) \nn
\label{CKM}
\eeq
Here
\[
V_{us} = \lambda,\quad V_{cb}=A\lambda^2,\quad 
V_{ub}=A\lambda^3(\rho -i\eta),\quad
V_{td} = A\lambda^3(1- \bar{\rho} -i \bar{\eta}) .
\]
While $|V_{cb}|$ and $|V_{ub}|$ can be extracted from
semileptonic $B$ decays, $|V_{td}|$ is at present probed in
the process of \bdbdbar\ mixing. 

The expression for the effective Hamiltonian 
describing $\Delta B = 2$ transitions is known at
next-to-leading order (NLO) in QCD perturbation theory of the Standard
Model~\cite{Buras}
\begin{eqnarray}
\label{hamilt}
H_{\mbox{eff}}^{\triangle B = 2} &=& \frac{G_F^2M_W^2}{4 \pi^2}
\left({V_{tb}}^{*}V_{td}\right)^2 \eta_B S_0(x_t)
\times\left[\alpha_s^{(5)}(\mu)\right]^{-6/23} 
\left[1+\frac{\alpha_s^{(5)}(\mu)}{4 \pi} J_5 \right] {\cal O}(\mu) 
\nonumber
\end{eqnarray}
where $G_F$ is a Fermi constant, $M_W$ is the W-boson mass, 
$\eta_B=0.55\pm0.1$~\cite{Buras:1990fn}, $J_5=1.627$ in the
naive dimensional regularization (NDR) 
scheme, $S_0(x_t)$ is the Inami-Lim function~\cite{Inami:1980fz}, and
${\cal O}(\mu)=(\bar b_L\gamma_{\sigma}d_L)(\bar b_L\gamma_{\sigma}d_L)(\mu)$
is a local four-quark operator at the normalization point~$\mu$. 
Mass splitting of heavy and light mass eigenstates is
\begin{eqnarray}
\label{offdiag}
\Delta m =2 |\langle\bar B^0|H_{\mbox{eff}}^{\triangle B = 2}|B^0
\rangle| 
=~
\!\! {\cal C}\left[\alpha_s^{(5)}(\mu)\right]^{-6/23} 
\left[1+\frac{\alpha_s^{(5)}(\mu)}{4 \pi} J_5 \right] 
\langle\bar B^0|{\cal O}(\mu)|B^0\rangle \nonumber
\end{eqnarray}
where we introduced a constant ${\cal C}=G_F^2M_W^2 
\left({V_{tb}}^{*}V_{td}\right)^2 \eta_B m_B S_0(x_t)/\left(4 \pi^2\right)$. 
The largest uncertainty in calculation of the mass splitting  
is introduced by the hadronic matrix element
${\cal A} = \langle\bar B^0|{\cal O}(\mu)|B^0\rangle$ 
that is poorly known~\cite{PDG}. 
The evaluation of this matrix element is a genuine non-perturbative
task, which should 
be approached with some non-direct techniques. The simplest approach 
(``factorization'')~\cite{Gaillard:1974hs} 
reduces the matrix element ${\cal A}$ to the product of simpler matrix
elements measured in leptonic $B$ decays
\be
{\cal A}^{f} = \frac{8}{3}
\langle\bar B^0|\bar b_L\gamma_{\sigma}d_L|0\rangle
\langle 0|\bar b_L \gamma^{\sigma}d_L|B^0\rangle 
= \frac{2}{3} f_B^2 m_B^2
\ee
where the decay constant $f_B$ is defined by
$
\langle 0|\bar b_L \gamma_\mu d_L|B^0({\bf p})\rangle = i p_\mu f_{B}/2
$ and $m_B$ is the $B^0$ meson mass.
A deviation from the factorization ansatz is usually described by the parameter
$B_B$ defined as
$
{\cal A} = B_B {\cal A}^{f}
$;
in factorization $B_B=1$.
The evaluation of this parameter (and the
analogous parameter $B_K$ of $K^0 - \bar K^0$ mixing) has long
history. Many different results were obtained within approaches based
on quark models, unitarity, ChPT. The approach of direct numerical
evaluation on the lattice has also been used.
The corresponding results can be found in the 
literature~\cite{Bardeen,ope-three-kk,bb-three,Reinders,Narison,Melikhov,lattice,Hiorth}.

In my talk I report on the results 
of the calculation of the hadronic mixing matrix elements 
using Operator Product Expansion (OPE) and QCD 
sum rule techniques for three-point 
functions~\cite{main1,ope-three-kk,bb-three,Reinders,svz,fesr}. 
This approach is very close in spirit to lattice computations~\cite{lattice}, 
which is a model-independent, first-principles method. 
The difference is that the QCD sum rule
approach uses an asymptotic expansions of a Green's function
computed analytically 
while on the lattice the function itself can be numerically computed
provided the accuracy of the technique is sufficient. 
The sum rule techniques also provide a consistent way of taking 
into account perturbative corrections to matrix 
elements which is needed to restore the RG invariance of
physical observables usually violated in the factorization
approximation~\cite{kkaplhas}.

To start with let me introduce the three-point correlation function
\begin{eqnarray}
\label{threepoint1}
\Pi(p_1,p_2)
=\int dx dy 
\langle 0|T J_{\bar B}(x) {\cal O}(0) \bar J_{B}(y)|0\rangle  
e^{i p_2 x - i p_1 y}
\end{eqnarray}
of the relevant $\Delta B=2$ operator ${\cal O}(\mu)$
and interpolating currents for the $B^0$-meson
$J_{B} = (m_b+m_d)\bar d i\gamma_5 b$. Here 
$m_b$ is the $b$ quark mass. The current $J_{B}$ is RG invariant and
$
J_{B} =\partial_\mu (\bar d \gamma_\mu\gamma_5 b)
$.
The main relevant property of this current is 
$
\langle 0|J_{B}(0)|B^0(p)\rangle =  f_{B} m_B^2
$
where $m_B$ is the $B$-meson mass.
A dispersive representation of the correlator reads
\be 
\label{dispdouble}
\Pi(p_1, p_2) \equiv \Pi(p_1^2, p_2^2, q^2)=
\int\frac{\rho(s_1, s_2, q^2)ds_1 ds_2}{(s_1-p_1^2)(s_2-p_2^2)}
\ee
where $q=p_2-p_1$. For the analysis of $B^0 - \bar B^0$ mixing within
the sum rule framework this correlator 
can be computed at $q^2=0$. 

Phenomenologically the matrix element $\langle\bar B^0| {\cal O}(\mu) |B^0\rangle$
determines the contribution of the $B$-mesons in the form of a double
pole to the three-point 
correlator
\begin{eqnarray}
\label{phenrepr}
\Pi(p_1^2, p_2^2, q^2)=
\frac{\langle J_{\bar B}|\bar B^0\rangle}{m_{B}^2-p_1^2}
\langle\bar B^0|{\cal O}(\mu)|B^0\rangle
\frac{\langle B^0|\bar J_{B}\rangle}{m_{B}^2-p_2^2}+\ldots
\end{eqnarray}
Because of technical difficulties of calculation, a practical way of 
extracting the $B^0 - \bar B^0$ matrix element
is to analyze the moments of the correlation function at 
${p_1^2=p_2^2=0}$ at the point $q^2=0$
\be 
\label{momentsdef}
M(i,j)\equiv 
\frac{\partial^{i+j}\Pi(p_1^2, p_2^2, 0)}
{i!j!\partial p_1^{2i} \partial p_2^{2j}}
=\int \frac{\rho(s_1, s_2, 0)ds_1 ds_2}{s_1^{i+1} s_2^{j+1}}\, .\nn
\ee
A theoretical computation of these moments reduces to
an evaluation of single scale vacuum diagrams 
and
can be done analytically with available tools
for the automatic computation of multi-loop diagrams. 
Note that masses of light quarks are small (e.g.~\cite{lightmasses})
and can be accounted for as small perturbation which is relevant for 
the problem of $B_s$ meson mixing~\cite{narison2}.
\begin{figure}[ht]
\begin{center}
\includegraphics[scale=0.7]{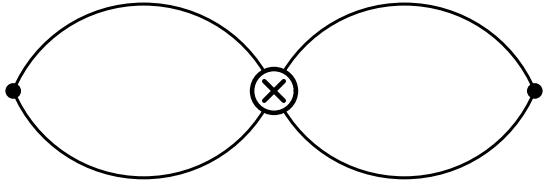}
\end{center}\caption{Perturbation theory diagram at LO}
\label{figLO}
\end{figure}
The leading contribution to the asymptotic expansion is given by 
the diagram shown in Fig.~\ref{figLO}. At the leading order in 
QCD perturbation theory the three-point function $\Pi(p_1, p_2)$
of Eq.~(\ref{threepoint1}) completely factorizes
\be
\Pi(p_1, p_2) = \frac{8}{3}\Pi_\mu (p_1)\Pi^\mu (p_2)
\ee
into a product of the two-point correlators $\Pi_\mu (p)$
\be 
\label{twopointscorr}
 \Pi_\mu (p)=p_\mu \Pi(p^2)=\int dx e^{i p x}
\langle 0|TJ_{\bar B}(x) \bar b_L \gamma_\mu d_L(0)|0\rangle .
\end{eqnarray}
At LO the calculation of moments is straightforward since
the double spectral density $\rho(s_1, s_2, q^2)$ is explicitly known
in this approximation. Indeed, using dispersion relation for the
two-point correlator 
\be 
\Pi(p^2) =\int_{m^2}^{\infty} \frac{\rho(s)ds}{s-p^2}, \quad
\rho(s)=\frac{3}{16\pi^2}m^2\left(1-\frac{m^2}{s}\right)^2
\ee
one obtains the LO double spectral density
in a factorized form
\be
\rho^{\rm LO}(s_1,s_2,q^2) =
\frac{8}{3} (p_1 \cdot p_2)\rho(s_1)\rho(s_2)=
\frac{4}{3} (s_1+ s_2-q^2)\rho(s_1)\rho(s_2)\ .
\ee
Thus, all PT contributions are of the factorizable form at LO.
First non-factorizable contributions to Eq.~(\ref{dispdouble}) appear
at NLO. Of course, at NLO there are also the factorizable diagrams.
Note that the classification of diagrams in terms of their
factorizability is consistent as both classes
are independently gauge and RG invariant. 

\begin{figure}[t]
\begin{center}
\includegraphics[scale=0.7]{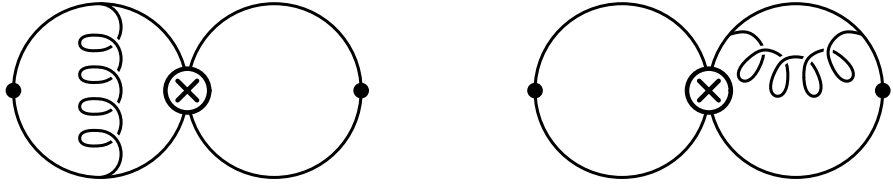}
\end{center}
\caption{Factorizable diagrams at NLO}
\label{figNLOfac}
\end{figure}

Consider first the NLO factorizable contributions 
that are given by the product of two-point 
correlation functions from Eq.~(\ref{twopointscorr}), 
as shown in Fig.~\ref{figNLOfac}.
Analytical expression for such contributions can be obtained as
follows. Writing 
$
\Pi(p^2)=\Pi_{\rm LO}(p^2)+\Pi_{\rm NLO}(p^2)
$
one finds
\be
\Pi_{\rm NLO}^f(p_1, p_2) = \frac{8}{3}(p_1.p_2)
(\Pi_{\rm LO}(p_1^2)\Pi_{\rm NLO}(p_2^2)
+\Pi_{\rm NLO}(p_1^2)\Pi_{\rm LO}(p_2^2)).
\ee
Since the spectral density of the correlator
$\Pi_{\rm NLO}(p^2)$ is known analytically
the problem of the NLO analysis in 
factorization is completely solved. Even a NNLO analysis of factorizable 
diagrams is possible as several moments of two-point correlators are 
known analytically. Others can be obtained numerically 
from the approximate spectral density~\cite{chetmomnondiag}. 

\begin{figure}[t]
\begin{center}
\includegraphics[scale=0.7]{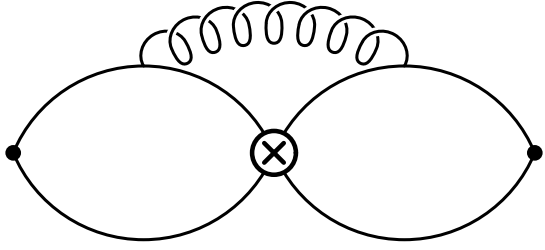}
\end{center}
\caption{A typical non-factorizable diagram at NLO}
\label{figNLOnonfac}
\end{figure}

The NLO analysis of non-factorizable contributions within perturbation
theory is the main point of 
my talk. The analysis amounts to the calculation of a set of 
three-loop diagrams (a typical diagram is presented in Fig.~\ref{figNLOnonfac}). 
These diagrams have been computed using the package MATAD for automatic calculation
of Feynman diagrams~\cite{matad}. The package is applicable only for
computation of scalar integrals.
the decomposition of the three-point amplitude
into scalars is known~\cite{Davyd}. 
A scalar function $A(p_1,p_2)$ of two four-momenta $p_1$ and $p_2$
can be expanded in a series over the scalar variable 
$p_1^2$, $p_2^2$ and $p_1\cdot p_2$ of the general form
\begin{eqnarray}
A(p_1,p_2) = \sum_{j_1,j_2,j_3} a_{j_1,j_2,j_3} 
({p_1^2})^{j_1}({p_2^2})^{j_2}(p_1\cdot p_2)^{j_3} .\nn
\end{eqnarray}
For the coefficients $a_{j_1,j_2,j_3}$ one finds
\begin{eqnarray} 
a_{j_1,j_2,j_3}=&&
\frac{1}{2^{j_1}2^{j_2}j_3!}
\sum_{k=0}^{[j_1+\frac{j_3}{2}]}\frac{(-1)^{k-j_1}}{2^{2k+j_2-j_1}(2j_1+j_3-2k)!}\nn\\
&&\sum_{l=max(0,k-j_1)}^{min(k,k+j_2-j_1)}
\frac{(2j_1+j_3-2k+2l)!}{l!(j_1-k+l)!}  \times \nonumber \\
&& \frac{
(1-\ep)_{2j_1+j_3-2k+2l+1}(1-\ep)_{j_1+j_3-k+l}}
{(1-\ep)_{2j_1+j_3-2k+2l}(1-2\ep)_{2j_1+j_3-2k+2l+1}}\nn\\
&&
 \frac{(1-\ep)_{2j_1+j_3-2k+l}(1-\ep)_{1}(1-2\ep)_{1}
}{
(1-\ep)_{2j_1+j_3-k+l+1}(1-\ep)_{j_1+j_2+j_3-k+l+1}
} \nonumber \\
&& \quad \times
\triangle_{11}^k\triangle_{22}^{k+j_2-j_1}
\triangle_{12}^{2j_1+j_3-2k}
A(p_1,p_2)|_{p_1=p_2=0}\nn
\end{eqnarray}
with
\[
\triangle_{ij} = g^{\mu\nu}
\frac{\partial^2}{\partial p_i^{\mu}p_j^{\nu}} .
\]
Here $(a)_b$ stands for the Pochhammer symbol 
$(a)_b = \Gamma(a+b)/\Gamma(a)$ 
and $d=4-2\ep$ is the spacetime dimension.
The actual calculation has been performed with the
computer algebra system FORM~\cite{form}. 

Renormalization of the local four-quark operator ${\cal O}$ entering the 
effective Hamiltonian has been done in 
dimensional regularization with an anticommuting $\gamma_5$.
The renormalization of the operator ${\cal O}$ reads
\be
{\cal O}^R={\cal O}^B-\frac{\alpha_s}{4\pi}\frac{1}{\ep}{\cal O}_c
\ee
with
$
{\cal O}_c=(\bar b_L \Gamma_{\mu\nu\alpha} t^a d_L)
(\bar b_L \Gamma^{\mu\nu\alpha} t^a d_L)
$. The color space matrices $t^a$ are the $SU_c(3)$ generators 
and 
$
\Gamma_{\mu\nu\alpha}=(\gamma_\mu\gamma_\nu\gamma_\alpha-
\gamma_\alpha\gamma_\nu\gamma_\mu)/2$.
Note that in four dimensional space-time the matrix 
$\Gamma_{\mu\nu\alpha}$ reduces to the expression
$-i \epsilon_{\mu\nu\alpha\beta}\gamma^\beta \gamma_5$.
This relation is however ill defined in $D$-dimensional spacetime of
dimensional regularization.
The renormalization of the factorizable contributions reduces 
to that of the $b$-quark mass $m$. 
We use the quark pole mass as a mass 
parameter of the calculation.

The expression for the ``theoretical'' moments reads
\be
\label{thfull}
M_{th}(i,j)=\frac{m^6 a_{ij}}{m^{2(i+j)}}
\left(1+\frac{\alpha_s}{4\pi} \left(b^{f}_{ij}+b^{nf}_{ij}\right)\right)
\ee
where the quantities $a_{ij}$, $b^{f}_{ij}$ and $b^{nf}_{ij}$
represent LO, NLO factorizable and NLO nonfactorizable contributions 
as shown in Figs.~\ref{figLO}-\ref{figNLOnonfac}. 
The NLO nonfactorizable contributions $b^{nf}_{ij}$ with 
$i+j\leq 7$ are analytically calculated in ref.~\cite{main1}
for the first time.
The calculation required about 24 hours of computing time on a dual-CPU 
2 GHz Intel Xeon machine. The calculation of higher moments is feasible 
but requires considerable optimization of the code. This work
is in progress.
The analytical result for
the lowest finite moment $M_{th}(2,2)$ reads
\be
a_{22}=\frac{1}{(16\pi^2)^2}
\left(\frac{8}{3}\right), \quad
b^{f}_{22}=\frac{40}{3}+\frac{16\pi^2}{9}\, ,
\ee
\be
b^{nf}_{22}
=
S_2\frac{8366187}{17500}-\zeta_3\frac{84608}{875}
-\pi^2\frac{33197}{52500}
-\frac{426319}{315000}\nn .
\ee
Here
$
S_2=\frac{4}{9\sqrt{3}}{\rm Cl}_2
\left(\frac{\pi}{3}\right)=0.2604\ldots
$, $\zeta_3=\zeta(3)$, and $\mu^2=m^2$. 
Higher moments contain the same transcendental entries 
$S_2$, $\zeta_3$, $\pi^2$ with different
numerical coefficients. The numerical values 
for the moments are
$
b^{nf}_{ij}
$:
$
b^{nf}_{2(2345)}=\{0.68,1.22,1.44,1.56\}
$
and 
$
b^{nf}_{3(34)}=\{1.96,2.25\}
$.
The above theoretical results are used to 
extract the non-perturbative parameter $B_B$ from the sum rules 
analysis.

\begin{figure}[t]
\begin{center}
\includegraphics[scale=0.7]{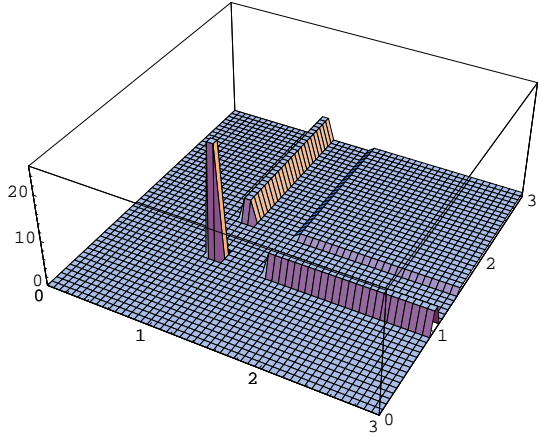}
\end{center}
\caption{A model of phenomenological spectrum}
\label{phenplot}
\end{figure}

The ``phenomenological'' side of the sum rules is given by the moments
which can be inferred from Eq.~(\ref{phenrepr})
\begin{eqnarray}
\label{phenfull}
M_{ph}(i,j)
= \frac{8}{3} B_B
\frac{f_B^4 m_B^2 }{m_B^{2(i+j)}}
+{\cal O}\left(\frac{1}{(m_B^2+\Delta)^{i+1}m_B^{2j}},
\frac{1}{(m_B^2+\Delta)^{j+1}m_B^{2i}}\right)\nn
\end{eqnarray}
where the contribution of the $B$-meson is displayed
explicitly. The remaining parts are the contributions due 
to higher resonances and 
the continuum which are suppressed due to the mass gap $\Delta$ in the 
spectrum model. A rough picture of the phenomenological spectrum is
given in Fig.~(\ref{phenplot}).

For comparison we consider the factorizable approximation
for both ``theoretical''
\be
\label{thfact}
M_{th}^{f}(i,j)=\frac{m^6 a_{ij} }{m^{2(i+j)}}\left(1
+\frac{\alpha_s}{4\pi} b^{f}_{ij}\right)
\ee
and ``phenomenological'' moments, which, by construction,
are built from the moments of the two-point function
of Eq.~(\ref{twopointscorr})
\be
\label{phenfactprod}
M_{ph}^{f}(i,j)
= \frac{8}{3}\frac{f_B^4 m_B^2}{m_B^{2(i+j)}}+...
\end{eqnarray}
According to standard QCD sum rule technique, the 
``theoretical'' calculation is dual to the ``phenomenological''
one. Thus, Eq.~(\ref{phenfull}) should be equivalent 
(in the sum rule sense) to Eq.~(\ref{thfull}). Also, in factorization,
Eq.~(\ref{phenfactprod}) is equivalent  to Eq.~(\ref{thfact}). 
Now Eq.~(\ref{thfact}) and
Eq.~(\ref{thfull}) differ only due to non-factorizable corrections. Therefore, 
the difference between Eq.~(\ref{phenfactprod}) and Eq.~(\ref{phenfull}) is
because the residues differ from their factorized values.

To find the nonfactorizable addition to $B_B$ from the sum rules 
we form ratios of the total and factorizable contributions.
On the ``theoretical'' side one finds
\be
\label{fintheory}
\frac{M_{th}(i,j)}{M_{th}^{f}(i,j)}
=1+\frac{\alpha_s}{4\pi}\frac{b^{nf}_{ij}}{1+\frac{\alpha_s}{4\pi}
b^{f}_{ij}}\, .
\ee
This ratio is mass-independent.
On the ``phenomenological'' side we have
\be
\frac{M_{ph}(i,j)}{M_{ph}^{f}(i,j)}=
\frac{B_B+R_B (z^j+z^i)+C_Bz^{i+j}}
{1+R^f(z^j+z^i)+C^f z^{i+j}}
\ee
where $z=m_B^2/(m_B^2+\Delta)$ is a parameter that describes the
suppression of higher state contributions. $\Delta$ is a gap 
between the squared masses of the $B$-meson and higher states. 
$R_B$, $C_B$, $R^f$ and $C^f$ 
are parameters of the model for higher state contributions within the sum
rule approach. 
In order to extract the non-factorizable
contribution to $B_B$ we write $B_B=1+\Delta B$.
Similarly, one can parameterize contributions to ``phenomenological'' moments 
due to higher $B$-meson states by writing $R_B=R^f+\Delta R$ and 
$C_B=C^f+\Delta C$. Clearly, $\Delta B=\Delta R=\Delta C=0$ in factorization. 
We obtain
\be
\label{finphen}
\frac{M_{ph}(i,j)}{M_{ph}^{f}(i,j)}=1
+\frac{\Delta B+\Delta R (z^j+z^i)+\Delta C z^{i+j}}
{1+R^f(z^j+z^i)+C^f z^{i+j}}\, .
\ee
Comparing Eqs. (\ref{fintheory}) and (\ref{finphen}) one sees 
how the perturbative non-factorizable correction $b^{nf}_{ij}$
is ``distributed'' 
among the phenomenological parameters of the spectrum. 
We extract $\Delta B$
by a combined fit of several ``theoretical'' and ``phenomenological'' moments.
The final formula for the determination of $\Delta B$ reads
\be
\frac{\alpha_s}{4\pi} b^{nf}_{ij}
=\Delta B
+\Delta R (z^{j-2}+z^{i-2})
+\Delta C z^{i+j-4}
\ee
where $\Delta R$ and $\Delta C$ are free parameters of the fit.
We take $\Delta=0.4 m_B^2$ for the $B$ meson 
two-point correlator. This corresponds to the duality interval of
$1~{\rm GeV}$ in energy scale for the analysis based on finite energy
sum rules~\cite{locduality}. The actual value of $\Delta B$
has been extracted using the least-square fit of all available
moments.
Estimating all uncertainties we
finally find the NLO non-factorizable QCD corrections to
$\Delta B$ due to perturbative contributions to the sum rules to be
\[
\Delta B=(6\pm 1)\frac{\alpha_s(m)}{4\pi}
\]
We checked the stability of the sum rules which lead 
to a prediction of $\Delta B$.
For $m_b=4.8~\GeV$ and $\alpha_s(m_b)=0.2$~\cite{Penin:1999kx} 
one finds $\Delta B=0.1$.
The calculation can be further improved with the evaluation of 
higher moments. The result is sensitive to the parameter $z$ or to
the magnitude of the mass gap $\Delta$ used in the parametrization of the
spectrum. 

In conclusion, the $B^0 -\bar B^0$ mixing matrix element 
has been evaluated in the framework of QCD sum rules for three-point functions at NLO
in perturbative QCD. The effect of radiative corrections on $B_B$ is under 
complete control within pQCD and amounts to approximately $+10$\% of
the factorized value.

\end{document}